\documentclass[conference]{IEEEtran}

\usepackage{booktabs} 
\usepackage{amsmath}
\usepackage{graphicx}
\usepackage{comment}

\usepackage{cite}
\usepackage{xcolor}
\usepackage{multirow}
\usepackage{pbox}
 
\usepackage{flushend}

\usepackage{subfigure}

\begin{document}

\title{Effectiveness
of Data Augmentation in Cellular-based Localization  Using Deep Learning}

\author{\IEEEauthorblockN{Hamada~Rizk}
\IEEEauthorblockA{\textit{Dept. of Comp. Sci. and Eng.,} \\
\textit{Egypt-Japan Univ. of Sci. \& Tech. (E-JUST) }\\
Alexandria, Egypt\\
hamada.rizk@ejust.edu.eg}
\and
\IEEEauthorblockN{Ahmed~Shokry}
\IEEEauthorblockA{\textit{Dept. of Comp. and Sys. Eng.,} \\
\textit{Alexandria University}\\
Alexandria, Egypt \\
ahmed.shokry@alexu.edu.eg}
\and
\IEEEauthorblockN{Moustafa Youssef}
\IEEEauthorblockA{\textit{Dept. of Comp. and Sys. Eng.,} \\
\textit{Alexandria University}\\
Alexandria, Egypt \\
moustafa@alexu.edu.eg}}

\maketitle

\begin{abstract}
Recently, deep learning-based positioning systems have gained attention due to their higher performance relative to traditional methods. However, obtaining the expected performance of deep learning-based systems requires large amounts of data to train model. Obtaining this data is usually a tedious process which hinders the utilization of such deep learning approaches.

In this paper, we introduce a number of techniques for addressing the data collection problem for deep learning-based cellular localization systems. The basic idea is to generate synthetic data that reflects the typical pattern of the wireless data as observed from a small collected dataset.

Evaluation of the proposed data augmentation techniques using different Android phones in a cellular localization case study shows that we can enhance the performance of the localization systems in both indoor and outdoor scenarios  by 157\% and 50.5\%, respectively. This highlights the promise of the proposed techniques for enabling deep learning-based localization systems.
\end{abstract}

\section{Introduction}

Location-based services for indoor and outdoor environments have become an essential part for our daily life.
 Generally, GPS is considered the de facto standard for ubiquitous and accurate outdoor navigation. 
However, it  requires line-of-sight to the satellites. Thus, it neither works well in urban regions nor indoors. In addition, it drains the phone battery quickly.
  As an alternative, several localization techniques have been proposed  which can be utilized in both indoor and outdoor scenarios. For example, some systems depend on the inertial sensors in smart phones to obtain the location \cite{alzantot2012uptime,alzantot2012crowdinside,abdelnasser2016semanticslam,aly2014map++,elbakly2018truestory,ibrahim2018cnn,mohssen2014s,elhamshary2016transitlabel,rizk2015hybrid,elhamshary2014checkinside,aly2015lanequest}. However, these low-cost sensors are usually noisy, leading to a quick degradation in accuracy. 
  On the other hand, WiFi-based techniques \cite{youssef2005horus,shokry2017tale,youssef2006location,kosba2012robust,abbaswideep,abdel2013monophy,elbakly2016robust,youssef2005multivariate,youssef2003wlan,youssef2003small,youssef2004continuous} are the most common in this domain due to the wide-deployment of WiFi APs. Nonetheless, all the above mentioned techniques require a high-end phone equipped with GPS, sensors, and/or WiFi. This limits their ubiquitous adoption, especially in developing countries. 
  
 On the contrary, since all phones support cellular technology by definition, it becomes desirable to use cellular signals for localization. Furthermore, this  consumes virtually no extra power in addition to the normal phone operation.
 A number of cellular-based localization techniques have been proposed in literature for both indoor \cite{rizk2018cellindeep,tian2015robust,varshavsky2007gsm,rizk2019Increasing} and outdoor settings \cite{ibrahim2012cellsense,ibrahim2010cellsense,ibrahim2011hidden,shokry2018deeploc,aly2013dejavu,crescendo}. 
 Nevertheless, current cellular-based positioning systems either try to learn the  pattern of the received signal strength using probabilistic techniques, e.g. \cite{ibrahim2010cellsense,ibrahim2011hidden,ibrahim2012cellsense}; or using traditional classifiers, e.g. SVM  \cite{tian2015robust} or KNN \cite{varshavsky2007gsm}. 
 These techniques  cannot learn a good mapping between signals and locations due to 
 their limited learning ability in the presence of the high noise inherent in the wireless channel. In addition, probabilistic approaches usually take the simplifying assumption that  the signals from the different APs are independent, limiting their accuracy.
To address these limitations, a number of localization systems  \cite{rizk2018cellindeep,shokry2018deeploc,wang2015deepfi,abbaswideep} have adopted deep learning. However, 
deep learning models are data-hungry; i.e. in order to achieve the expected good performance they require large amounts of training data. This comes with extra costs due to the need to perform the tedious data collection process  over a longer time. 
 
 In this paper, we introduce a new data collection framework for  cellular-based localization systems that use deep neural networks (DNN). The proposed framework can be integrated seamlessly with any of the current localization systems by processing their small input dataset to produce a larger dataset suitable for training deep learning models.
In particular, we introduce different data augmentation techniques that can automatically provide large amount of data for the training process starting from a \textbf{small set} of collected data. The proposed techniques:  the sampling, additive noise, cell tower dropper, and deep generative augmenter; are based on analyzing the behavior of the wireless channel and generating synthetic data that reflects this behavior. Note that data augmentation techniques have been used before in image~
 \cite{krizhevsky2012imagenet,lecun1998gradient}
and audio~\cite{
schluter2015exploring} recognition systems. However, these techniques are usually domain-specific and cannot be directly extended to cellular-based localization systems.

  We have implemented and deployed the proposed framework and data augmentation methods on different Android phones and evaluated its performance  in indoor and outdoor testbeds in a typical deep learning localization system.  Our results show that the proposed data augmentation techniques achieve a consistent median accuracy of 0.77m and 87m in indoor and outdoor scenarios, respectively. This reflects an improvement of more than  157\% and 50.5\% in the indoor and outdoor performance compared to the case of no data augmentation. 
  
  The rest of the paper is organized as follows:   The proposed framework is presented and  the details of the different methods of data augmentation is described in Section ~\ref{sec:system_details}.
  We evaluate the proposed data augmentation  framework in indoor and outdoor scenarios in Section \ref{sec:evaluation}.
Finally, Section \ref{sec:conclusion} concludes the paper.

\section{Proposed Data Augmentation Framework}
\label{sec:system_details}
\begin{figure}
\centering
\includegraphics[width=0.99\linewidth,height=5.3cm]{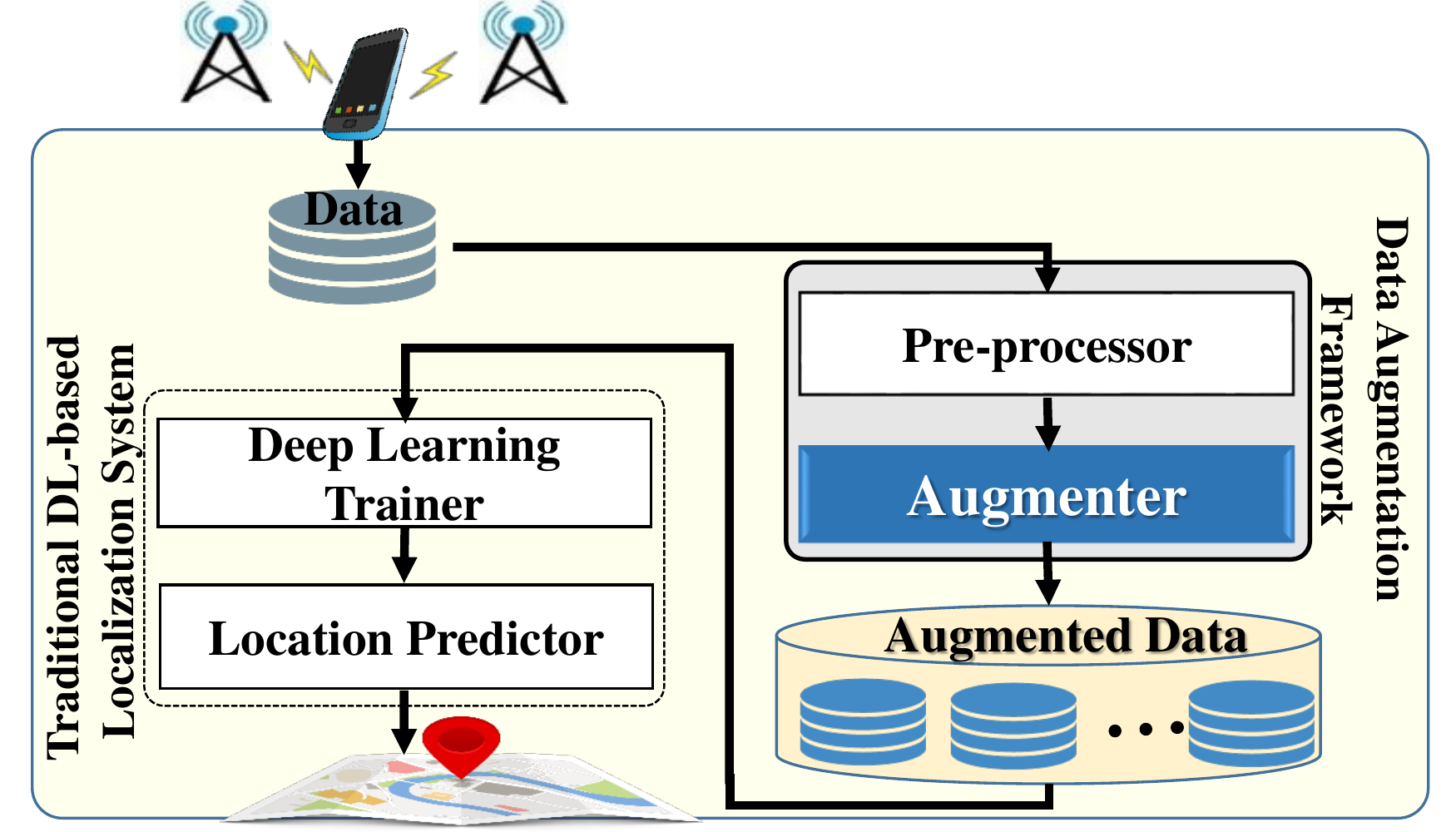}
\caption{The proposed data augmentation framework. The framework works on the small input dataset of the traditional localization system to generate a larger data set suitable for training deep learning models.}
\label{fig:framework}
\end{figure}

Fig.~\ref{fig:framework} shows how our data augmentation framework can be integrated with any of the current deep learning-based cellular localization systems \footnote{Without loss of generality, we focus in this paper on cellular-based localization systems due to their ubiquity and low-energy consumption. However, our proposed data augmentation techniques can be applied to any other deep learning-based RF localization systems such as WiFi-based systems.}. Basically, the framework takes as input the data collected by the traditional localization system, typically \textbf{small in size},  and generates a new \textbf{larger} dataset that should be used for training the localization system. Our framework contains two sub-modules: the Pre-Processor and the Data Augmenter. In the balance of this section, we start by describing the data collection process and input data format followed by the details of the Pre-Processor and the Data Augmenter modules. The later is the core contribution of this paper.

\subsection{Data Collection}

To collect the training data, the cell phone scans for the cell towers at different reference locations i.e. points (indoors) or  grid cells (outdoors) in the area of interest.
According to the standard \cite{ibrahim2012cellsense}, up to seven towers of the total
$m$ available cell towers in the area can be heard at any scan. 
For each heard cell tower, a pair (CID, RSS) is recorded, where CID represents the cell tower unique ID and RSS is the
received signal strength from that tower. Different RSS scans are performed at each reference location to collect the training data.

\subsection{The Pre-Processor Module}
The goal of the \textbf{Preprocessor} is to normalize and transform the scanned RSS vectors from the heard cell towers in the area of interest to fit the format required for the deep network training.
Theoretically, RSS in cellular networks is expressed in decibels (dB) with reference to one milliwatt (mW), i.e. dBm units. However, we leveraged the Android cellular API to obtain the cell readings, which reports the RSS in the Arbitrary Strength Unit (ASU). The ASU is an integer value ranging from 0 to 31 and is proportional to the RSS measured in dBm based on the following relation:
\begin{equation}
dBm = 2 \times ASU -113
\end{equation}
To improve the convergence time of the deep learning model, we normalize these ASU values directly, to the range [0,1].

The feature vector per sample is constructed from all cell towers detectable in the area of interest. Since not all these towers can be overheard in every scan, this module assigns zero to the towers that are not heard in a given scan.
This allows us to fix the feature vector size that is input to the neural network.

\subsection{The Data Augmenter}
\label{sec:aug}

In this section, we present different data augmentation techniques that can be used to extend a small input dataset of cellular scans to a larger dataset suitable for training deep learning models. In particular, we describe the details of the additive noise, sampling, cell tower dropper, and deep generative techniques.

\subsubsection{Additive noise}
Typically, when both the transmitter and receiver are stationary, the RSS changes over time. This is due to the dynamic changes in the environment and the noisy wireless channel. 
The additive noise data augmentation technique builds on this observation. In particular, for each input cellular scan vector $x$, it adds a white Gaussian noise $q$ with a specific standard deviation $s$ to each entry. The standard deviation varies for different cell towers for different fingerprinting locations. More formally, 
\begin{equation}
q \sim \mathcal{N} (0,s^2)
\end{equation}
and
\begin{equation}
\tilde x_{ij} \sim  x_{ij} + \mathcal{N}
(0, s_{ij}^2) = \mathcal{N}(x_{ij}, s_{ij}^2)
\end{equation}

Where $x_{ij}$ is the original RSS measurement for each cell tower $j$ at each location $i$ and $s$ is the corresponding standard deviation taken proportional to the signal range at a specific location as:
\begin{equation}
s_{ij} =  \frac{max(x_{ij})-min(x_{ij})}{2}
\end{equation}

\begin{figure}[!tbp]
\tiny
  \centering
   \begin{minipage}[b]{0.89\linewidth}
  \includegraphics[width=1.0\linewidth,height=4.9cm]{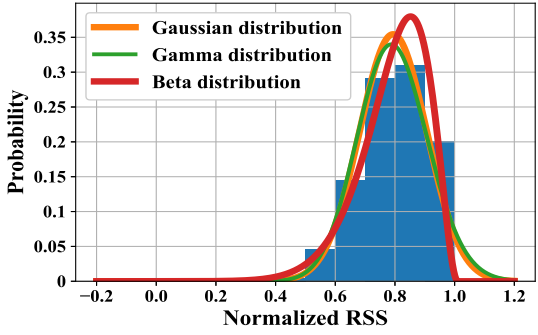}
\caption{Fitting different distributions to the cell tower \textbf{normalized} RSS at a certain location. The Beta distribution provides the best fit.}
\label{fig:fitting_distribution}
\end{minipage}  
\end{figure}

\subsubsection{Sampling technique} \label{sec:sampling}
 This technique constructs the signal strength distribution from the small input dataset. Then it generates synthetic data by sampling from this distribution. 
 
 To find the optimal distribution that fits the input data of one cell tower at a particular fingerprint location, we use maximum likelihood estimation over the parameters of different common distributions. 
We found that  the Beta, Gamma and Gaussian distributions are the ones that provide the best fit 
for most cell towers at different locations. Specifically, the Beta distribution generally yields the best likelihood value, followed closely by the Gamma distribution, followed by the Gaussian distribution (Fig.~\ref{fig:fitting_distribution}). This can be explained by noting that the Pre-Processor module normalizes the range of the input RSS to be between 0 and 1. This fits the range of the Beta distribution more than the infinite range of the Gamma and Gaussian distributions.

To generate a sample using this technique, a random value is drawn from the distribution corresponding to the  $j^{th}$ tower at the $i^{th}$ location. Subsequently, synthetic values of different towers are combined together to form a complete sample at some location.

\subsubsection{Tower dropping techniques}
\begin{figure}[!tbp]
\tiny
  \centering
  \begin{minipage}[b]{0.80\linewidth}
\includegraphics[width=1.0\linewidth,height=4.5cm]{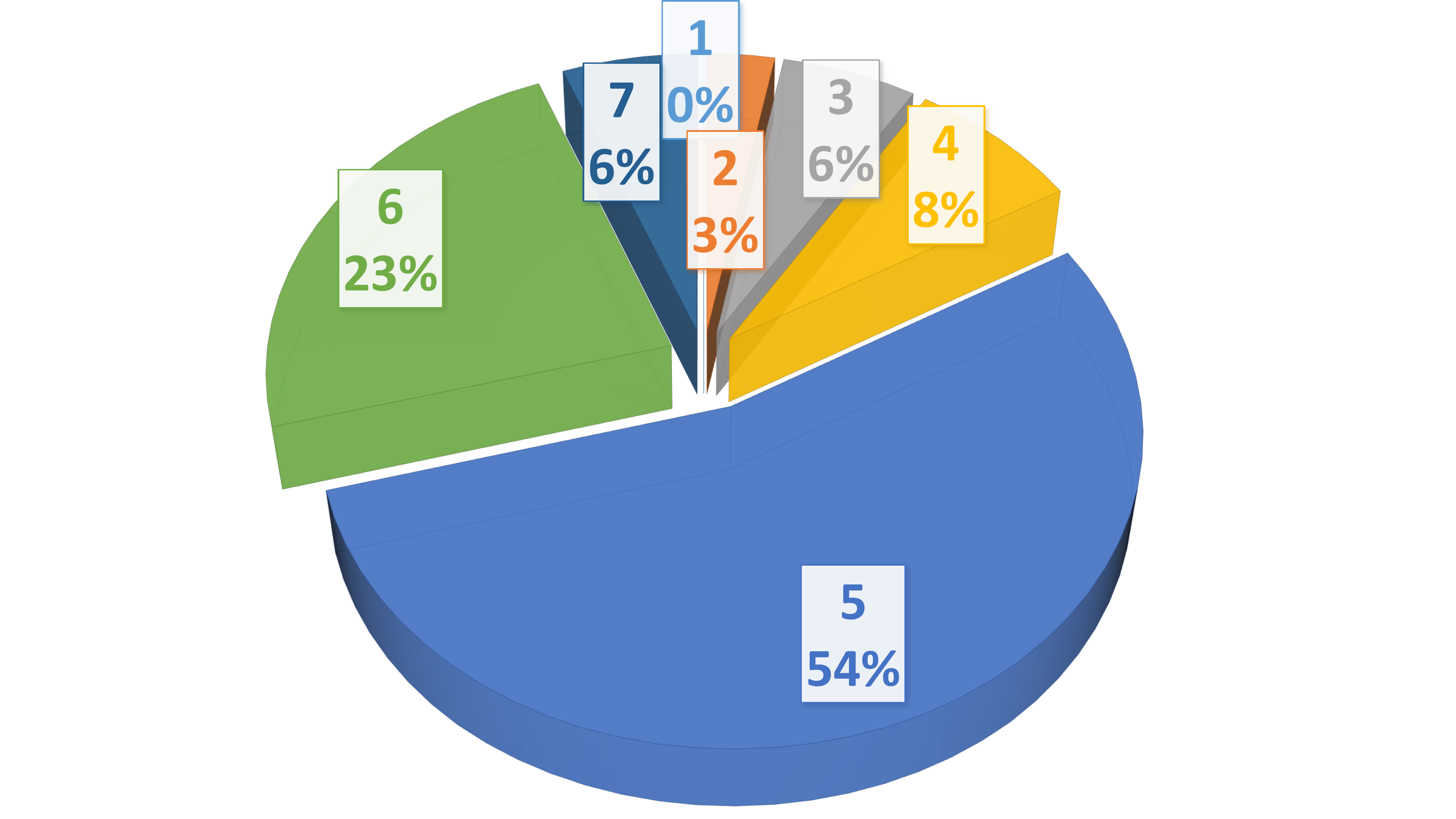}
\caption{Probability of hearing
different number of cell
towers at the same location (indoor).}
\label{fig:tower_density}
  \end{minipage}
\end{figure}

When the device is stationary at a certain location, the number of cell towers it hears over different scans varies as shown in Fig.~\ref{fig:tower_density}. The figure shows that even though the phone is able to detect seven cell towers per scan, it detects only up to 5 towers 71\% of the time. This is due to the receiver's  sensitivity\footnote{The receiver sensitivity refers to the minimum signal strength the wireless chip is able to detect a signal at.}, where weak cell towers, whose signal changes over time, will be dropped randomly.

\begin{figure}
\centering
\includegraphics[width=0.95\linewidth,height=6.0cm]{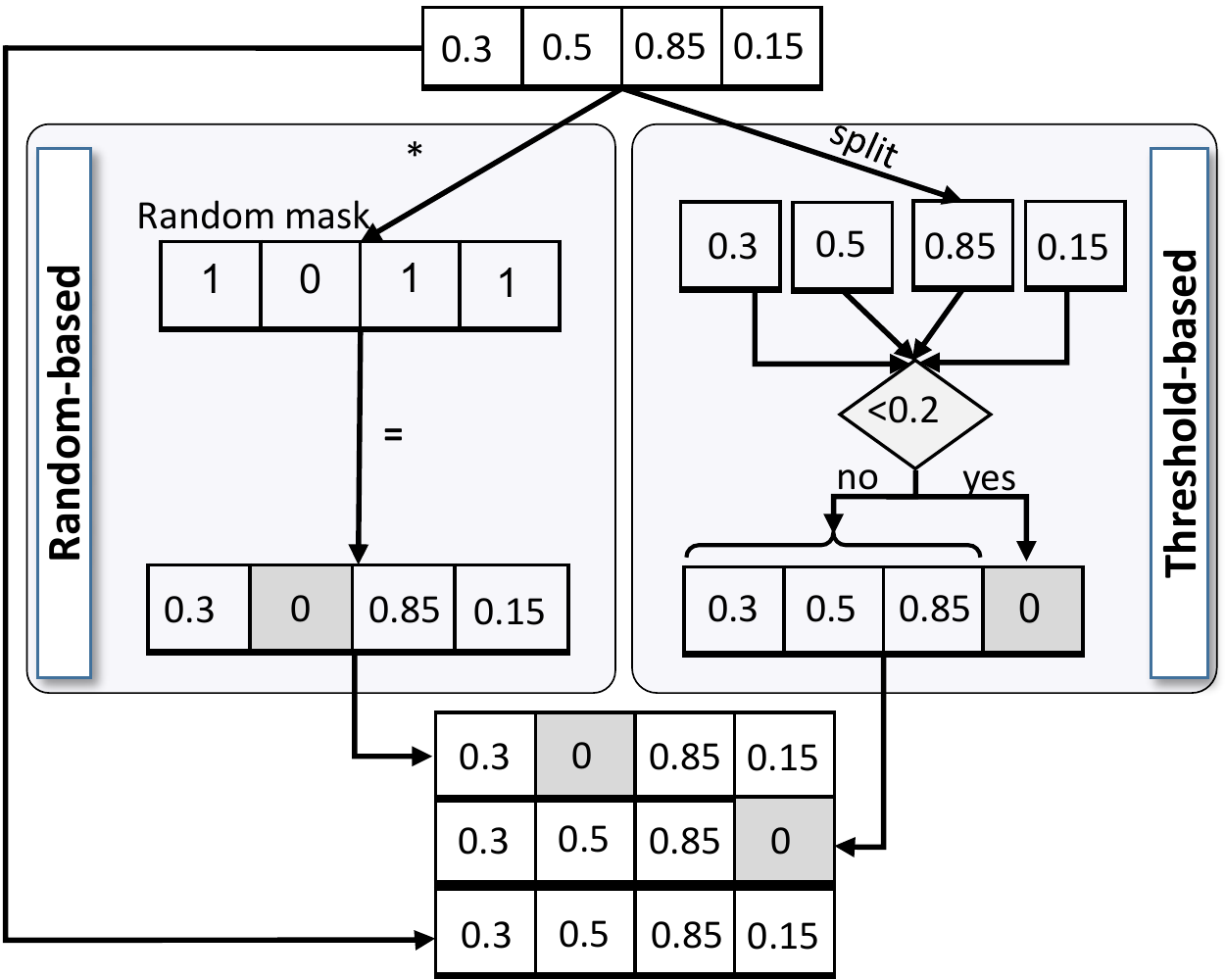}
\caption{Cell tower dropper techniques employ dropping a number of cell towers RSS to emulate the noisy channels. The normalized scan vector is multiplied by random masks to obtain the transformed vector out of the  \textbf{Random} method. On the other hand, the RSS values that are less than a threshold (e.g. $0.2$ ) are dropped in The   \textbf{Threshold} method. 
}
\label{fig:dropper}
\end{figure}

The tower dropping techniques increase the training data size by generating synthetic training samples that reflect this behavior by randomly dropping cell towers (setting their values to 0) in the input scan.  We introduce two novel tower dropping techniques:
The random-based dropper and the threshold-based dropper.  
Fig.~\ref{fig:dropper} shows an example for the two techniques.

\textbf{Random-based dropper: }
The idea is to randomly drop (set RSS to 0 ASU) signals from the detected cell towers (other than the associated one to reflect the real case). This can be done by randomly generating a sparse binary vector that is then multiplied with  the original RSS vector. This operation masks out some RSS readings from the RSS vector, simulating the case when the different neighboring cell towers appear and disappear at the same location over time.

\textbf{Threshold-based dropper: }
This technique modifies an input RSS vector such that any entry $s_i$  whose value falls below a predefined threshold is a candidate to be removed (set to zero), mimicking that this cell tower has not been heard in the generated synthetic sample. All combinations of removing these entries can then be added to the training dataset.

\subsubsection{Deep generative technique}

This technique is similar to the Sampling technique described in Section \ref{sec:sampling}. However, the sampling technique learns the distribution of each cell tower independent from the other cell towers while the deep generative technique learns the joint distribution between the different cell towers.  

In particular, this data augmenter employs variational autoencoder (VAE) models at different fingerprinting locations to learn the joint distribution $P(X_i)$ of input features over training samples $X_i$ at fingerprint point $i$. VAE are popular approaches to unsupervised learning of complicated distributions \cite{kingma2013auto,goodfellow2014generative}. As such, the technique generates new samples from $P(X_i)$ at reference location $i$.

\begin{figure}[!tbp]
\centering
\includegraphics[width=0.95\linewidth,height=5.0cm]{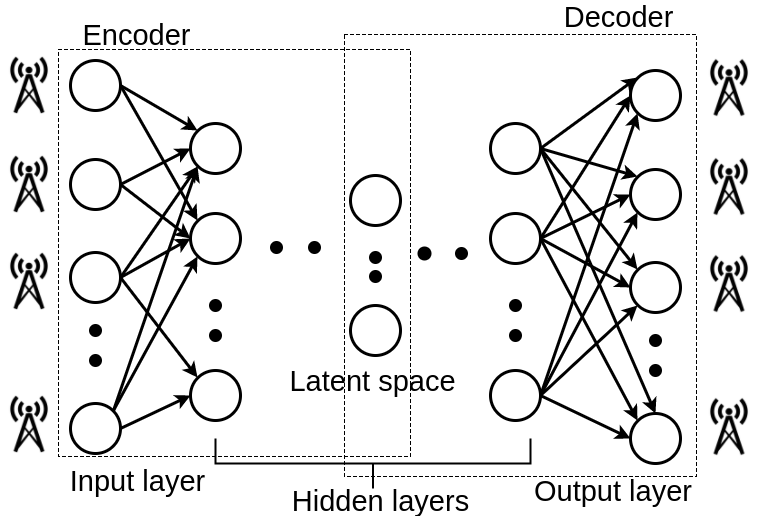}
     \caption{Variational Autoencoder network structure.}
     \label{fig:vae-model}
\end{figure}

Fig.~\ref{fig:vae-model} shows the VAE network structure. The VAE consists of an encoder, a decoder, and a loss function. The encoder is a neural network that compresses a data point input $x \in X_i$ to get the hidden (latent) representation $z$ and network weights and biases $\theta$ as an output. The latent space $z$ is typically referred to as a "bottleneck" because the encoder must learn an efficient compression of the data into a lower-dimensional space. 
We refer to the encoder as $q_{\theta}(z|x)$.
We can sample from this distribution to get noisy values of the representations $z$. The decoder is another neural network that takes the latent representation $z$ as an input and produces the parameters to the probability distribution of the data and has weights and biases $\phi$. The decoder is denoted by $p_{\phi}(x|z)$. The loss function for sample $x_i$ is the negative log-likelihood function defined as follows,
\begin{equation}
    L_i(\theta, \phi) = - E_{z\sim q_\theta(z\vert x)}[\log p_\phi(x\vert z)] + KL(q_\theta(z\vert x) \vert\vert p(z))
\end{equation}

The loss consists of two terms. First, the reconstruction loss. It is the expected negative log-likelihood of the $i$-th data point. This term forces the decoder to learn to reconstruct the data. The second term is the regularizer. It is the Kullback-Leibler \cite{joyce2011kullback} divergence between the encoder's distribution $q_\theta(z\vert x_i)$ and a predefined distribution $p(z)$. Assuming that $p(z) = N(0,1)$, the regularizer forces the latent representations $z$ to follow the standard normal distribution. We employ the stochastic gradient descent to optimize the loss with respect to network parameters $\theta$ and $\phi$. For training the VAE deep model, We tried different network structures. The best structure in terms of training accuracy contains three hidden layers with 10, 5 (latent space), and 10 neurons. Network training is performed over 3000 epochs at a learning rate of 0.001. 

After finishing the training phase, we can generate new data points by sampling the latent variables $z \sim N(0,1)$, then decode $z$ to get a new data point $\hat{x}$.

\section{Evaluation}
\label{sec:evaluation}
In this section, we evaluate the performance of the proposed techniques in two different scenarios: an indoor and an outdoor testbed. We start by describing the experimental setup used in the data collection process.
Next, we analyze the effect of different data augmentation techniques in each scenario on the localizaiton systems' performance. 

\subsection{Experimental setup} \label{sec:expSetup}

\begin{figure}[!t]
    \centering
    \subfigure[Indoor testbed.]
    {
        \includegraphics[width=.46\linewidth,height=3.5cm]{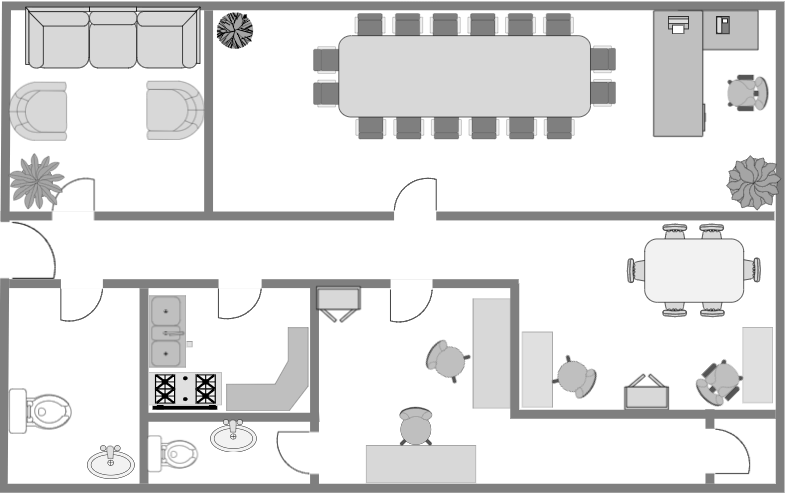}
        \label{fig:floorplan}
    }
    \subfigure[Outdoor testbed.]
    {
      	\includegraphics[width=.46\linewidth,height=3.5cm]{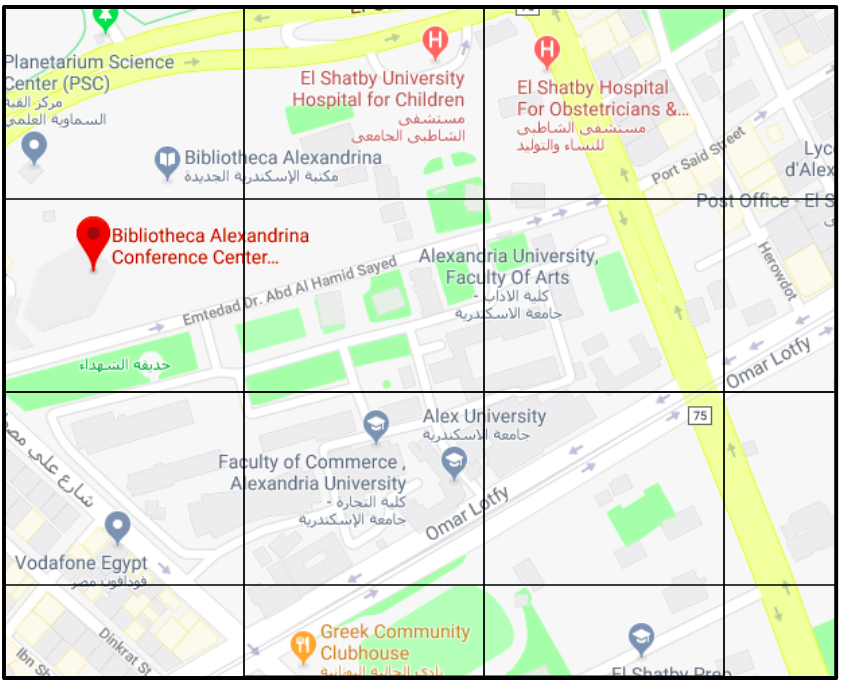}	\label{fig:outdoor-map}
    }
    \caption{Indoor and outdoor testbeds.}
	\label{fig:cdfs}
\end{figure}

To collect the necessary data for evaluation, we deployed our system in two different scenarios. The first one is an indoor testbed  with a $11m\times12m$ area containing offices, meeting rooms as well as corridors (Fig.~\ref{fig:floorplan}). Data is collected at uniformally distributed reference points that span the area of interest with a total of 55 points. This is done while the user is stationary  over the course of two minutes. The ground-truth location is set by the user on the floorplan during the data collection process.
The second testbed (outdoor) covers  0.2K$m^2$ in an urban area. The area of interest is partitioned virtually into a grid of equally-sized square cells (as shown in Fig.~\ref{fig:outdoor-map}) \cite{shokry2018deeploc,ibrahim2012cellsense}. Cellular data is collected by war-drivers and each cell is considered as one location. The war drivers' devices collect the GPS locations, which we use as the ground-truth locations. 

In both testbeds, we use different Android devices including HTC Nexus One, Prestigio Multipad Wize 3037 3G, HTC One X9 and Motorola Moto G5 Plus phones among others. 

 We implemented the data collector App using the Android SDK to scan cell towers. The program records the (cell identifier, RSS, timestamp) for each heard tower in the area of interest. The scanning rate was set to one scan per second. 
Without a loss of generality, we adopt the localization systems in \cite{rizk2018cellindeep,shokry2018deeploc} with their reported optimal parameters as shown in Table \ref{table:parameters}. These systems employ  a multinomial (multi-class)  classifier
as a deep model for localization.
In the two systems, the input to the classifier is the RSS coming from cell towers that cover the entire area of interest.
The output is the reference locations probability distribution, i.e. the probability that the input
RSS scan belongs to each reference location (class) in the area of interest.
Finally, the estimated location is the average of all fingerprint reference locations weighted by their corresponding probability.

\begin{table}[!t]
            \centering
            \caption{Testbeds parameters.}
            \label{table:testbed}
            {\renewcommand{\arraystretch}{1.3}
            \begin{tabular}{|l|l|l|}
            \hline
            \textbf{Criteria} & \textbf{Indoor} & \textbf{outdoor}  \\ \hline 
             \textit{ Area }     &  11$\times$12 $m^2$ & 0.2 $\text{Km}^2$\\\hline 
              \textit{ Grid cell length } & 1 $m$ & 100 $m$\\\hline 
              \textit{ Total number of cell towers } & 17 & 37\\\hline 
              \textit{ Sampling rate (scan/sec) } & 1 scan/sec & 1 scan/sec\\ \hline 
            \end{tabular}
            }
        \end{table}

\subsection{Localization accuracy}
In this section, we evaluate the location accuracy of the proposed framework and the effect of data augmentation methods, described in Section~\ref{sec:aug}, on the indoor and the outdoor testbeds. Table~\ref{summary-results} summarizes the results.

\subsubsection{Indoor scenario}
\begin{figure}[!tbp]
\centering
\includegraphics[width=0.99\linewidth,height=5.5cm]{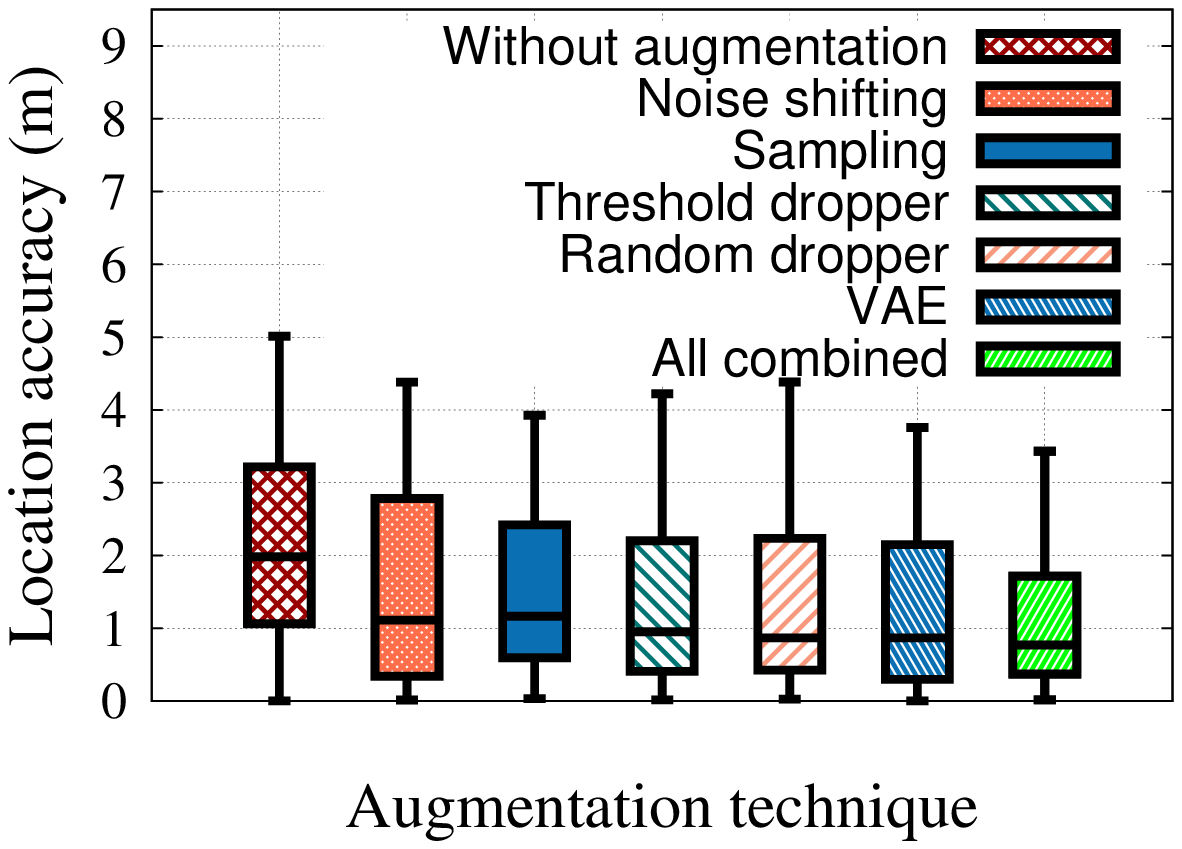}
     \caption{Effect of different data augmentation techniques on the indoor testbed accuracy.}
     \label{fig:indoorAccuracy}
\end{figure}

Fig.~\ref{fig:indoorAccuracy}  shows the performance of the proposed framework with different methods of data augmentation. 
The figure shows that all data augmentation techniques enhance accuracy compared to the case of not using augmentation at all. The VAE surpasses all other techniques  due to its ability to capture the joint distribution between the different cell towers. 
Nonetheless, combining the data from all data augmenters leads to the best accuracy, with an enhancement of 157\% in median error. 
 This is due to the corresponding increase of the training samples in addition to the learning of different patterns of cellular signals that affect the wireless channel in practice, making the system more robust to noise.

\subsubsection{Outdoor scenario}
\begin{figure}[!tbp]
\centering
\includegraphics[width=0.99\linewidth,height=5.5cm]{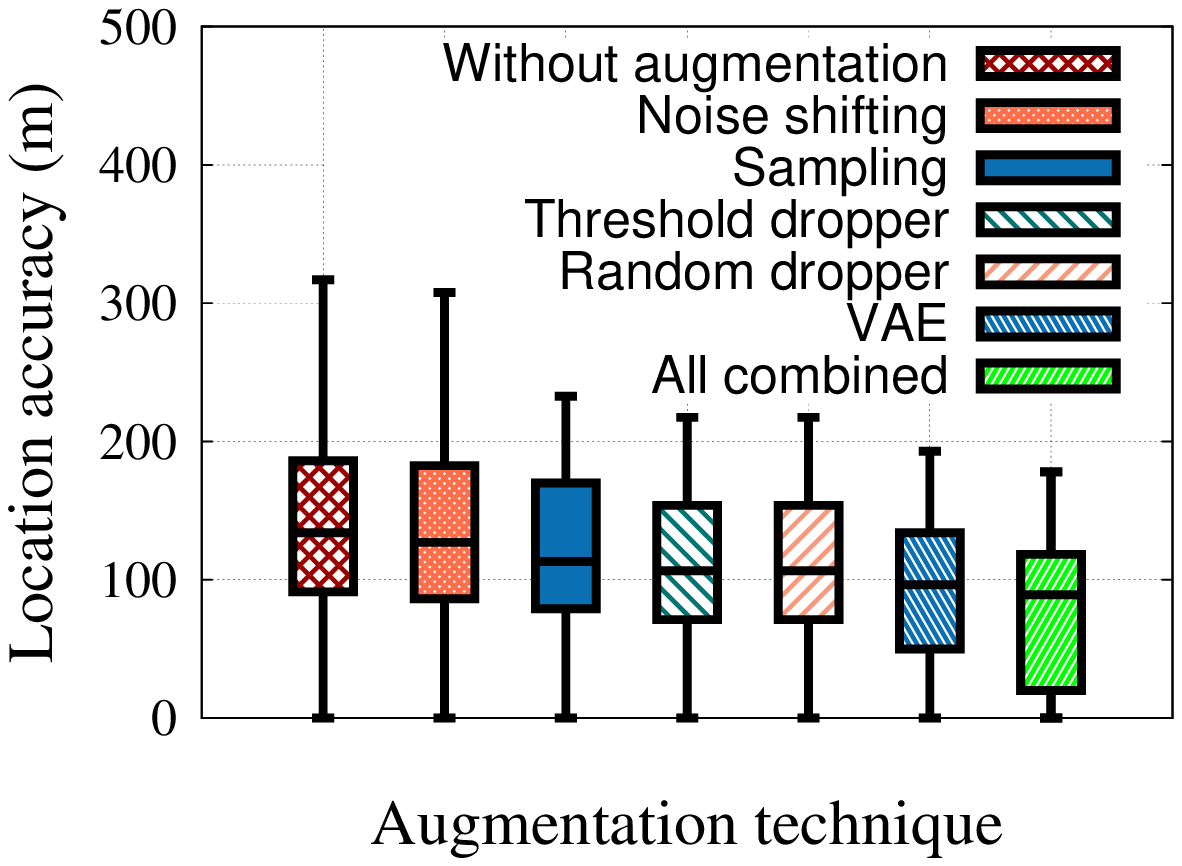}
     \caption{Effect of different data augmentation techniques on outdoor testbed accuracy.}
     \label{fig:aug_out}
\end{figure}

Fig.~\ref{fig:aug_out} shows the effect of different data augmentation techniques on the
localization
accuracy outdoors. 
The figure shows that, similar to the indoor case, data augmentation techniques can improve the system accuracy compared to training without augmentation. 
The figure also confirms that the VAE technique is superior to other techniques with a median localization accuracy of 97m. 
The figure also confirms that combining the data from all data augmentation techniques leads to the best accuracy with an enhancement of 50.5\% in median error compared to training without augmentation.

\begin{table}[!t]
            \centering
            \small
            \caption{Default hyper parameters for indoor and outdoor testbeds used in evaluation.}
            \label{table:parameters}
            \scalebox{0.9}{
            {\renewcommand{\arraystretch}{1.3}
            \begin{tabular}{|l|l|l|l|}
            \hline
             \textbf{Parameter} & \textbf{Indoor}  & \textbf{Outdoor} \\ \hline\hline
             \textit{ Learning rate }& 0.001 & 0.005 \\ \hline
             \textit{ Batch size }& 256 & 40 \\ \hline
             \textit{ Dropout rate (\%)}& 10 & 10 \\ \hline
             \textit{ Number of epochs }& 260 & 500 \\ \hline
             \textit{ Size of input  layer }& 17 & 37 \\ \hline
             \textit{ Number of hidden Neurons }& 280 & 345 \\ \hline
             \textit{ Number of hidden layers }& 4 & 3 \\ \hline
             \textit{ Size of output layer}& 55 & 20 \\   \hline
            \end{tabular}
            }
            }
    \end{table}

\section{Conclusion}
\label{sec:conclusion}
In this paper, we have investigated different data augmentation techniques for localization in cellular networks. We proposed a general framework for data augmentation that can be integrated with any deep learning fingerprinting-based  technique to increase its training dataset. 

We have implemented and evaluated the impact of each data augmentation technique on the model's localization accuracy. Our results show that the data augmentation techniques can significantly improve the localization accuracy by more than 157\% and 50.5\% for the indoor and outdoor testbeds, respectively. This justifies the use of data augmentation in deep cellular  localization systems.

\section*{Acknowledgment}
  This work is supported in part by a grant from the Egyptian National Telecommunication Regulatory Authority (NTRA).

\begin{table}[!tbp]
	\centering
	\caption{Summary for indoor and outdoor testbeds results.}
	\label{summary-results}
	\setlength\tabcolsep{4.3pt}
	{\renewcommand{\arraystretch}{1.3}
	\begin{tabular}{|l|l|l|l|l|}\hline
		Testbed & Accuracy & \pbox{10cm}{$25^{th}$\\percentile (m) }& \pbox{10cm}{$50^{th}$\\percentile (m)}&\pbox{10cm}{$75^{th}$\\percentile (m)}\\ \hline
		\multirow{2}{*}{Indoor} & \textbf{With aug.} &  \textbf{0.37}  & \textbf{0.77} & \textbf{1.71} \\ \cline{2-5}
		                        & Without aug. & 1.08 (-191.8\%) &  1.98 (-157\%) & 3.22 (-88.3\%) \\  \hline \hline                        
		\multirow{2}{*}{Outdoor} & \textbf{With aug.} &  \textbf{20}  & \textbf{89} & \textbf{118} \\ \cline{2-5}
		                        & Without aug. & 91 (-355\%) &  134 (-50.5\%) & 186 (-57.6\%) \\  \hline

	\end{tabular}
	}
\end{table}

\IEEEpeerreviewmaketitle

\ifCLASSOPTIONcaptionsoff
  \newpage
\fi

\end{document}